# Locational Marginal Price Variability at Distribution Level: A Regional Study

Calum Edmunds*, Waqquas A. Bukhsh, Simon Gill, Stuart Galloway
University of Strathclyde
Scotland, UK
*calum.edmunds@strath.ac.uk

*Abstract*— As distribution systems move towards being more actively managed there is increased potential for regional markets and the application of locational marginal prices (LMPs) to capture spatial variation in the marginal cost of electricity at distribution level. However, with this increased network visibility can come increased price volatility and uncertainty to investors. This paper studies the variation in LMPs in a section of the south west of England distribution network for current and future installed capacity of distributed generation. It has been shown that in an unconstrained network, spatial LMP variation (due to losses) is minimal compared to the temporal variation. In a constrained network, a significant increase in LMP volatility was observed, both spatially and temporally. This could bring risk for generators particularly if they become stranded in low price areas, or flexible demands facing a drop-off in return when constraints are removed.

## I. Introduction

When considering investment in distributed generation (DG) such as wind and solar the 'bankability' i.e. return on investment and associated risk, is a primary factor in a decision to invest. With decreasing subsidy support from the UK government for renewable generation, the long-term price uncertainty relating the revenues from DG becomes increasingly important [1]. At the same time with the unprecedented uptake in DG, anticipated to grow to up to 50% of installed capacity in the UK by 2050 [2], there is a growing need for price signals at distribution level to capture regional constraints and inform investment decisions. One option is the introduction of nodal pricing or locational marginal prices (LMPs) to distribution networks. The application of LMPs to distribution could offer advantages to a future distribution system operator (DSO) in managing network constraints and providing pricing reflective of losses and congestion [3]. However, a major concern with the application of LMPs is the volatility in prices and the effect this would have the bankability of revenue particularly for smaller distributed energy resource (DER) projects [4].

Previous work has been carried out in reducing nodal price variability on a small test network using demand side load shifting [5]. Using an electric equivalent network representative of the Spanish distribution system LMPs were calculated to be an average of 27 % higher at an LV node than at transmission [3]. Agent based modelling has been used to observe response of generators to variation in LMPs on a 5-bus test network [6], in the same work cross-correlation between LMPs in four neighbouring balancing authorities in the MISO region of the US was also carried out for 4 days in 2008. A literature review on the application of LMPs to distribution has been carried out [7] summarising work on electric vehicle charging [8] and markets with low voltage participants [9] amongst others. In terms of general price volatility, work using frequency domain analysis has been used to separate periodic price variations from random ones [10]. This included a locational study of 2000 nodes where 83 nodes exhibited significantly higher volatility, most likely due to constraints around these nodes. A large body of work has been conducted in electricity price forecasting, summarised in [11]. This is a widely researched area with complex methods such as multi-agent simulation and machine learning used to forecast prices (including price spikes) with reasonable accuracy.

This paper studies the variation in LMPs in a GB distribution network down to 11 kV for several cases including current and future installed capacity. It is aimed to compare LMP variation, both spatially and temporally, to a system wide wholesale price. This approach aims to quantify the additional risk to investors in DERs with the introduction of LMPs to distribution.

The layout of this paper is as follows; Section II presents a literature review of relevant LMP based research focussing on North America, where LMP based models are well established. Section III describes the methodology used in this paper including the network model and model input assumptions. Section IV outlines the results in terms of LMP variation. Finally, Sections V and VI provide discussion and conclusions from the work conducted.

This work has been supported through the EPSRC Centre for Doctoral Training in Wind and Marine Energy Systems (EP/G037728/1) and supported in part by EPSRC grant (EP/R002312/1)

## II. NODAL PRICING IN NORTH AMERICA

In established nodal day-ahead and real-time markets in North America - including NYISO, ERCOT, MISO, PJM and CAISO - the pricing and granularity can be so volatile that many participants choose not to participate directly in these markets [12]. LMPs are generally only applied to generation, while broad average zonal prices are applied to other agents [3]. In these markets most electricity is traded in monthly, annual, or yearly forward contracts to reduce exposure to risk from more volatile day ahead and real-time prices [13]. In the UK, where nodal pricing is not applied, the same effect is observed with 85% of electricity traded on forwards contracts in 2015 [14]. This suggests that without nodal pricing, there is sufficient risk from price volatility in day-ahead and real-time markets in the UK to encourage hedging that risk with forward trading; others have suggested that this is due to a poor spot market in the UK [13].

Volatility in nodal pricing in North American markets is usually hedged by the issuing of financial transmission rights (FTRs) issued by the system operator via forward auction. FTRs entitle the owner to revenues or charges for an agreed quantity of MW between two points on a network (source and sink nodes) [15]. The difference in LMPs at transmission level due to losses is generally between 5 – 10% [3], however, variations in LMPs of an order of magnitude are observed at points in time in US wholesale markets when network constraints occur. In US transmission systems, between 2009 and 2015, congestion has decreased overall due to lower demand, increased use of demand response and network upgrades [16]. In 2010, the average summer peak LMP across the MISO, PJM, New York and New England markets was below $50/MWh in the west to over $100/MWh in demand centres in the east [16]. More extreme price spikes can come at hourly resolution.

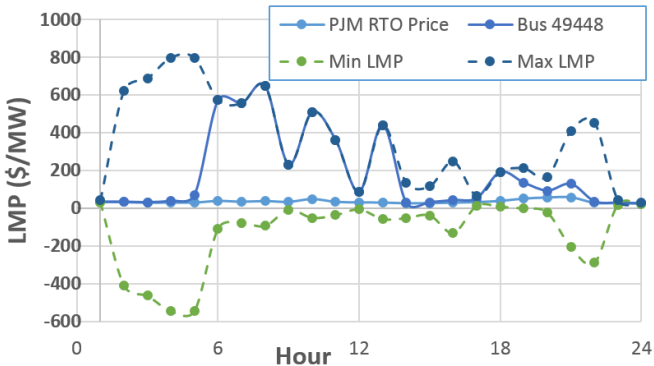

Figure 1 - PJM Prices: min, max, bus 49448 and RTO (average for PJM market), Tuesday 6th February 2018. Data from PJM Website [17]

Figure 1 shows a recent day when the average price for the PJM ranged from 25 to 58 $/MWh, while the minimum and maximum LMP at certain nodes ranged from 800 $/MWh to -546 $/MWh. This suggests that, at least in the PJM, the variation in LMP spatially is higher than that temporally.

## III. METHODOLOGY

### A. Network Model

A section of the south west (SW) of England transmission network (400 kV) has been modelled including associated 132 kV distribution network, all 33 kV primary substations, a single 33 kV network, along with an example 11 kV feeder. An overview of the 400 kV and 132 kV network is shown in Figure 2. Any references made to the SW England network made in this report refer to the network in Figure 2. This network was chosen due to containing large amounts of DG, particularly solar, with many areas reaching network capacity limits for new DG connections [18]. The boundaries nodes of the model (HINP and CHIC on the north east of the network) connect the SW England to the wider GB transmission network. These boundary nodes are modelled as generators that can absorb or generate power at a fixed price (assumed to be the wholesale GB market index price – MIP [19]). A distribution network down to 11 kV is added to the transmission system in south west part of the network (at node RAME1).

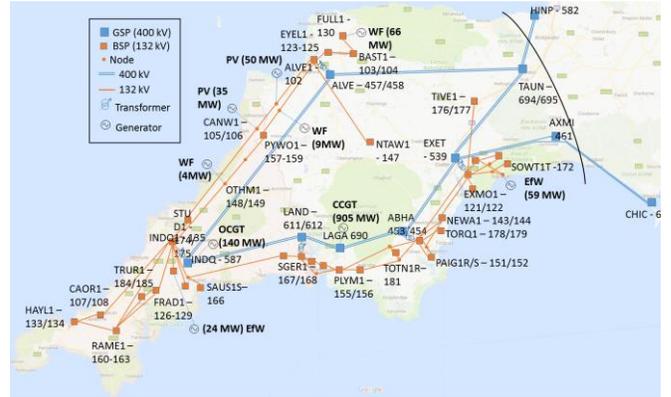

Figure 2 – Simplified 400 kV and 132 kV South West England Network Map

### B. Model inputs and assumptions

To calculate LMPs at each network node, an AC optimal power flow (OPF) was carried out using Matpower [1] on a 221 bus model of SW England. Model inputs (peak demands, generation capacities, branches, transformers, impedances etc.) for the 400 kV network were taken from the National Grid electricity ten year statement (ETYS) [20]. Data for the 132 kV network and below were taken from the distribution network operators (DNOs) long term development statement (LTDS) [21]. No 11 kV network information was available from the DNO therefore an 11 kV feeder from another region in the south of England is used in the model. Future generation capacity (Table I) has been estimated from connection applications (for renewable generation) made in the LTDS assuming all are accepted, even beyond network capacity (assuming non-firm connections). It is assumed that generation will be dispatchable such that it can be curtailed in the event of generation exceeding network capacity.

The normalised grouped demand of SW England grid supply points (GSPs) in 2015 is used as the demand profile.

Peak demand is assumed to be the same for the future case as for the base case, this is in line with the LTDS where peak demand changes very little in 5 years [21]. The generation profile from PV and wind generators are taken from the renewables output simulation tool developed by Staffell [28] and Pfenninger [29] for a general location in SW England. Electricity prices are the EPEX SPOT UK market index price (MIP) for 2015 [19]. The grid import/export points shown at the right-hand side of Figure 2 are fixed at the MIP. Prices for generation are assumed to be 150 £/MWh for OCGT plant, 50 £/MWh for CCGT plant and for all renewable generators price is assumed to be 0 £/MWh to reflect the short run cost.

TABLE I. GENERATION TYPES AND DEMAND WITH VOLTAGE LEVEL

|  | Voltage (kV) | | | | | | |
|---|---|---|---|---|---|---|---|
|  | Current Capacity (%) | | | Total (MW) | Future Capacity (%) | | Total (MW) |
| **Gen** | **400** | **132** | **33** |  | **33** | **11** |  |
| Gas | 100 | 0 | 0 | 1045 | 0 | 0 | 1045 |
| Wind | 0 | 45 | 55 | 264 | 67 | 0.8 | 362 |
| PV | 0 | 9 | 91 | 980 | 94 | 0.7 | 1601 |
| Biomass | 0 | 31 | 69 | 354 | 78 | 0.6 | 497 |
| **Total (%)** | 40 | 12 | 48 | 2642 | 61 | 0.5 | 3506 |
| **Peak Demand** | 300 | 0 | 1550 | 1861 | 1550 | 10.6 | 1861 |

## IV. RESULTS

To observe the variability of LMPs in a study distribution network now and in the future (assuming significant future increase in installed DG), several cases are considered:

1. *Base Case - current capacity (1 Day)*
   a. *Winters day – Maximum demand, Minimum DG*
   b. *Summers day – Minimum demand, Maximum DG*
2. *Future installed DG capacity (1 Day)*
   a. *Winters day*
   b. *Summers day*
3. *Time series (1 Year)*
   a. *Current capacity*
   b. *Future capacity*

For the base case and future capacity cases, results are shown over 1 day for the following buses;

- 107 – 132 kV bus close to large amounts of DG
- 244 – 33 kV bus with low installed DG
- 288 – 11 kV bus embedded in distribution with no DG
- 582 – 400 kV grid import/export bus

The timeseries results are shown for grid import/export bus 582 and 11 kV bus 288 over 1 year.

*1) Base Case - Current capacity.*

*a) Winters day – Maximum demand, Minimum DG*

In the base case, on both summer and winters days, no constraints are observed. Differences in LMPs between nodes are solely due to losses. On the winter's day (Figure 3), the SW network is importing from the external grid, therefore the highest LMPs are observed at the lowest voltages with resistive losses being highest at lower voltage. Bus 288, which is one of the most electrically remote 11 kV nodes, has the highest LMPs over this time period with LMPs ranging from 9% up to 16% above the grid import price at bus 582. It is worth noting that nodes at lower voltages will have the largest price spikes.

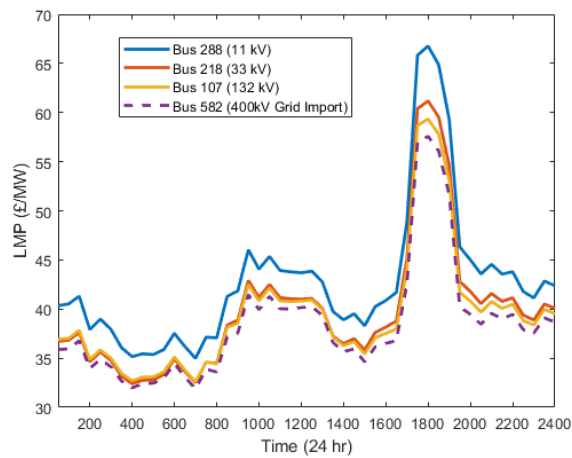

Figure 3 - LMP at buses 288, 218, 107 and 582 for Jan 18th current capacity

*b) Summers day – Minimum demand, Maximum DG*

During the summers day in June (Figure 4) PV output is above 50% between 11am and 6pm resulting in the SW exporting to the grid. During these times nodes with proximity to low cost renewable generation (e.g. buses 218) become the cheapest due to reduced losses. On summer days with peak generation and minimum demand, bus 218, which has a 4 MW PV array attached, will have prices lower than the grid import price by up to 14%.

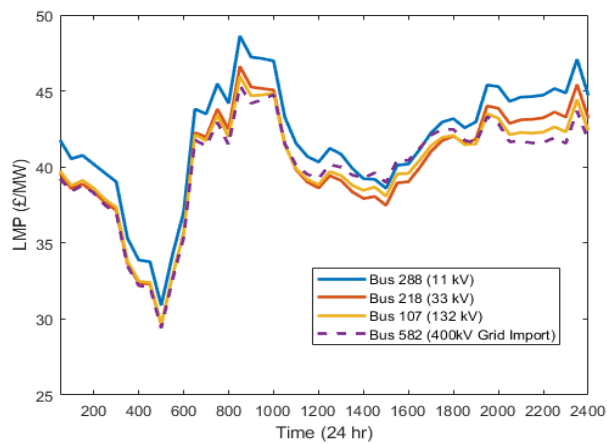

Figure 4 - LMP at buses 288, 218, 107 and 582 on June 5th current capacity

### 2) Future Installed Capacity

#### a) Winters day – Maximum demand, Minimum DG

The LMPs are very similar to the current capacity (Figure 3) however prices at all voltages are closer together due to a small increase in generation at lower voltages.

#### b) Summers day – Maximum demand, Minimum DG

In the future capacity case an additional 864 MW DG (predominantly PV) is connected, mainly aggregated to 33 kV primary substations. As PV output increases in the summer (Figure 5) congestion pricing is observed between midday and 5 pm. Due to the reverse power flow limit on 132/33 kV transformers, the output from PV in the 33 kV network is curtailed resulting in LMPs of 0 at 33 kV and 11 kV buses during these hours. This assumes DG has been allowed to connect beyond network capacity in non-firm connection agreements as is increasingly common in active network management schemes [22]. The prices at the 132 kV and 400 kV nodes are unaffected at these times as there is no constraint between these nodes and the wider network.

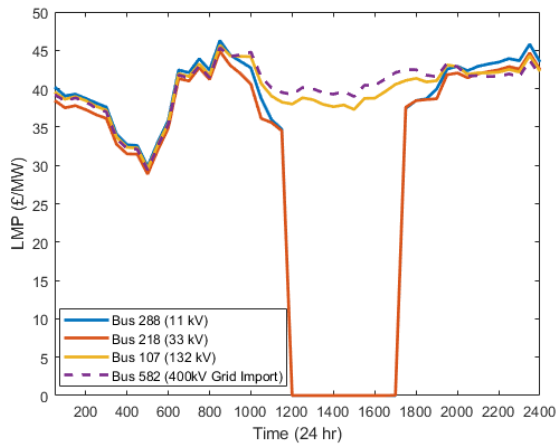

Figure 5 - LMP at buses 288, 218, 107 and 582 for June 5th future capacity

### 3) Time series

#### a) Current capacity

In Figure 6, with no constraints occurring over the year, the LMP of bus 288 follows the import price at bus 582. As the network is importing more often than not the average price of bus 288 is 8% higher than 582 due to losses.

Figure 7 shows the average LMPs over the year at current capacity. The lowest average prices are at 132 kV and 400 kV buses and highest at 11 kV.

Table II shows that the spatial variability (indicated by standard deviation) in LMPs at 400 kV is very low due to minimal losses, this increases with decreasing voltage level.

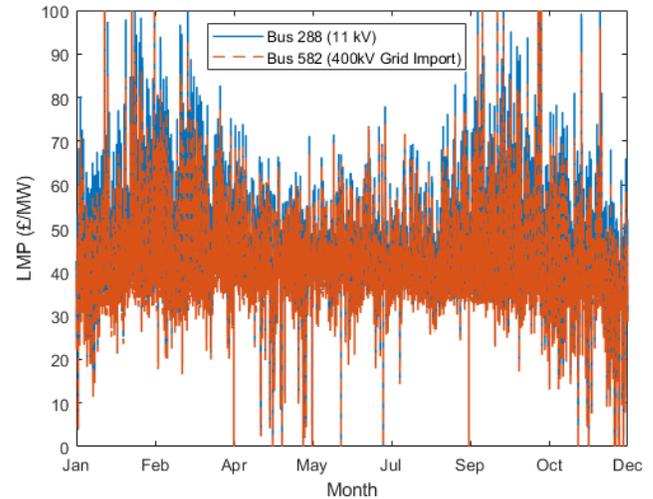

Figure 6 - LMP at 11 kV bus 288 and 400 kV grid import point bus 582 for current capacity

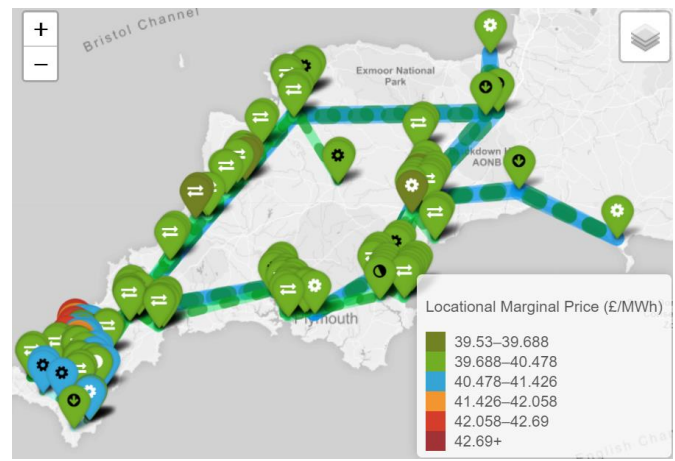

Figure 7 – Average LMP over SW England for current capacity

TABLE II.  LMP WITH VOLTAGE LEVEL – CURRENT CAPACITY

| Voltage level (kV) | LMP (£/MW) | | | Instances of 0 LMPs (%)[b] |
|---|---|---|---|---|
| | Mean | Spatial Std Dev[a] | Min/Max | |
| Grid Import | 39.92 | - | -32.3 / 259.2 | - |
| 400 | 40 | 0.08 | -25.2 / 259.4 | 0.0 |
| 132 | 40 | 0.24 | -25.6 / 264.6 | 0.0 |
| 33 (Rame) | 40.4 | 0.35 | -26.0 / 272.8 | 0.0 |
| 11 | 41.4 | 0.76 | -26.6 / 292.4 | 0.0 |

a. Average of half hourly standard deviation across nodes at each voltage
b. Number of instances of 0 LMPs as percentage of total. Not including instances with MIP <= 0.

#### b) Future capacity

In Figure 8 the LMP of bus 288 becomes much more volatile dropping to zero for 598 hours over the year. The LMPs drop to zero for buses behind a constraint where zero cost renewable generation is curtailed.

Figure 9 shows that the average LMP is lowest at 33 kV nodes in the far south west whereas transmission nodes have average LMPs up to 17% higher.

In Table III the mean LMP becomes lower at 33 kV and 11 kV than the MIP (as the region is exporting). The mean spatial LMP standard deviation is highest at 33 kV due to branch constraints within the network resulting in pockets of 0 LMPs at high DG output. The 11 kV spatial standard deviation is very similar to the current capacity case as all 11 kV nodes tend to show the same price fluctuations (e.g. dropping to 0) at the same time.

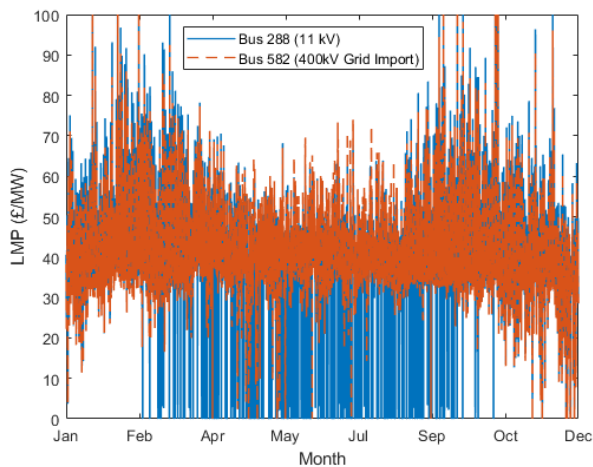

Figure 8 - LMP at 11 kV (bus 288) and 400 kV grid import point (bus 582) for future capacity (ACOPF)

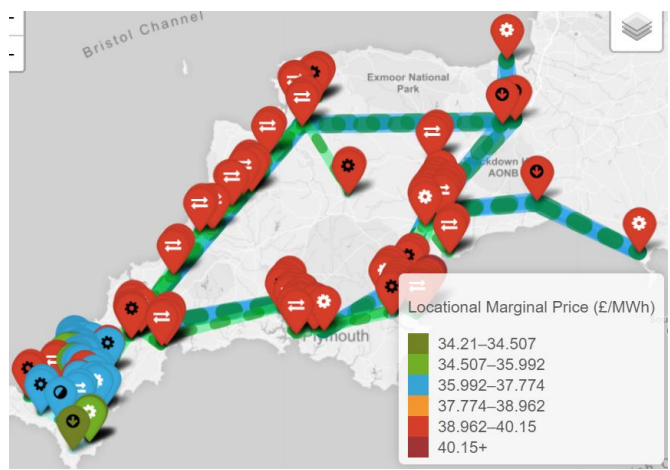

Figure 9 - Average LMP over SW England for future capacity

TABLE III. LMP WITH VOLTAGE LEVEL – FUTURE CAPACITY

| Voltage level (kV) | LMP (£/MW) | | | Instances of 0 LMPs (%)[b] |
| --- | --- | --- | --- | --- |
| | *Mean* | *Spatial Std Dev[a]* | *Min/Max* | |
| 400 | 39.9 | 0.09 | -32.6 / 259.3 | 0.0 |
| 132 | 39.8 | 0.25 | -33.2 / 263.3 | 0.0 |
| 33 (Rame) | 36.1 | 1.04 | -33.8 / 268.1 | 4.5 |
| 11 | 36.4 | 0.75 | -34.7 / 279.2 | 21.3 |

a. Average of half hourly standard deviation across nodes at each voltage
b. Number of instances of 0 LMPs as percentage of total. Not including instances with MIP <= 0.

## V. DISCUSSION

Price uncertainty is important for investors in estimating the return on any investment. Over the course of a day the GB MIP will generally follow a demand curve with prices higher at peak times and lower off-peak. Within this general trend there are many price spikes, Figure 6 shows the 2015 MIP (at bus 582) where prices exceed 90 £/MWh on 38 half hours. In the absence of network constraint, LMP variation spatially is predictable, increasing with distance from generation.

With current capacity, the application of LMPs will have a low impact in terms of LMP variability. Without any constraints occurring in the network, the LMPs at the 221 buses vary by an average standard deviation of 0.8 over a year. Comparing price variation temporally to spatially, the standard deviation for the MIP in 2015 was 10.53 over the whole year, 5.97 on Jan 18th and 3.46 on June the 5th. This shows that in an unconstrained network spatial variation with LMPs is minimal compared to over temporally.

When network constraints arise, there is increased volatility in LMPs, particularly in constrained areas, such as the 33 kV and 11 kV networks in the future capacity case where generation can exceed export capacity. If it was desirable to apply LMPs with less volatility they could be aggregated to 132 kV or 400 kV 'zones'. However, this risks losing visibility of constraints within these zones. For example, Figure 5 shows that LMPs at 132 kV mask constraints occurring at 33 kV and below.

Temporal price volatility is also increased when network constraints arise, particularly if a node is within a congestion zone, for example the standard deviation in LMP at 11 kV bus 288 over a year increases from 11.8 at current capacity to 15.6 for future capacity. Temporal variability is also high for 33 kV behind constraint, this reduces significantly at 132 kV which has temporal standard deviations close to that of the MIP. Again, aggregation of LMPs could be applied to reduce temporal volatility but at the cost of providing effective price signals to embedded generation or storage.

### A. LMP Investment risk

LMPs display a step change from constrained to unconstrained, for example in Figure 5 the LMP drops from £35/MWh to 0 in half hour 24 as soon as zero cost generation is curtailed. Returns will be greatly affected even with the addition of 1 MW (or indeed 0.1 MW) of generation, demand or network capacity if it tips the area in or out of constraint. Therefore it has been rightly suggested by others [3] that investors should be coordinated to avoid the risk of inefficient investment. A rush to build generation could rapidly lead to generation constraint and diminished returns for all (assuming non-firm connection agreements). Likewise, a method of properly allocating rewards could be considered (such as long-term contracts) for flexible demand to prevent free-riding where every user within a constraint zone benefits from an investment (i.e. lower congestion charging) without paying for it.

## B. Applying LMPs at distribution in the UK

In considering the application of LMPs at distribution in the UK, comparison can be made to the PJM, which serves a customer base of 65 million, over twice the 28 million domestic customers in the UK distribution system. The PJM uses LMPs calculated at transmission level for around 11,500 nodes by DCOPF. Losses are estimated using loss penalty factors for each node. The PJM bus model (available on their website) contains 7133 nodes at sub 138 kV which is often considered distribution and 3890 nodes at sub 35 kV which is well into the realms of distribution in the UK. The SW of England contains around 1.4 million electricity customers. There are around 2000 nodes in the DNOs published network data for the region which includes all 33 kV network but no detail around 11 kV network [21]. It is therefore possible to foresee an LMP based market applied to 33 kV level for each of the UKs 10 DNO regions, however to include all 11 kV networks which aren't currently published by DNOs (which suggests the data has not been collected), would require significant investment of time and money. In terms of computational power, for comparison the ERCOT market in Texas serves 24 million customers and requires thousands of servers to run the day-ahead and real-time optimisation.

## VI. CONCLUSIONS AND FUTURE WORK

The work presented in this paper has identified that, in the absence of constraints, the variation of LMPs due to losses is minimal on average compared to the temporal variation which GB wholesale electricity market participants are exposed to.

When constraints arise, with the growth in non-firm connections, LMPs can become much more volatile temporally and spatially. In the network in question, constraints were mainly observed at the 132/33 kV transformers, therefore LMP volatility was seen at 33 kV and 11 kV, not at 132 kV or 400 kV. Therefore, a zonal model aggregated to 132 kV or 400 kV would not effectively capture these constraints using LMPs. The average LMP is reduced in constrained areas where generation is curtailed, at 11 kV with regular constraints the price may be an average of 9% lower (in the case of high DG penetration at 33 kV and below) than the transmission LMP. This could benefit local flexible demand but would impact on the returns of generators in these areas until sufficient flexible demand is present to remove the constraint. Co-location of storage and generation would likely be most profitable to benefit from both scenarios; however, this would need to be regulated to prevent price fixing.

Limitations of this modelling are that short run costs of zero are assumed for all renewable generation. Future work could be in including agent modelling to simulate bidding behaviour and subsequent price fluctuations in times of constraint including negative pricing. To improve the accuracy of simulating network constraints, a SCOPF could be carried out preferably including some loss estimation.

Another aspect to be explored further is the addition of flexible demand for utilising constrained generation.

REFERENCES

[1] D. Newbery, "High level principles for guiding GB transmission charging and some of the practical problems of transition to an enduring regime," 2011.

[2] National Grid, "Future Energy Scenarios," 2017.

[3] Massachusetts Institute of Technology, "Utility of the future, An MIT Energy Initiative response to an industry in transition," 2016.

[4] K. Bell and S. Gill, "Delivering a highly distributed electricity system: Technical, regulatory and policy challenges," *Energy Policy*, vol. 113, no. November 2017, pp. 765–777, 2018.

[5] L. Goel, Q. Wu, and P. Wang, "Reliability enhancement and nodal price volatility reduction of restructured power systems with Stochastic demand side load shift," *2007 IEEE Power Eng. Soc. Gen. Meet. PES*, pp. 1–8, 2007.

[6] H. Li, J. Sun, and L. Tesfatsion, "Separation and Volatility of Locational Marginal Prices in Restructured Wholesale Power Markets," *Econ. Work. Pap.*, no. 138, pp. 1–34, 2010.

[7] C. Edmunds, S. Galloway, and S. Gill, "Distributed Electricity Markets and Distribution Locational Marginal Prices: A Review," in *Universities Power Engineering Conference (UPEC), 2017 52nd International*, 2017.

[8] R. Li, Q. Wu, and S. S. Oren, "Distribution locational marginal pricing for optimal electric vehicle charging management," *IEEE Trans. Power Syst.*, vol. 29, no. 1, pp. 203–211, 2014.

[9] E. Ntakou and M. Caramanis, "Price discovery in dynamic power markets with low-voltage distribution-network participants," *2014 IEEE PES T&D Conf. Expo.*, pp. 1–5, 2014.

[10] F. L. Alvarado and R. Rajaraman, "Understanding price volatility in electricity markets," *Proc. 33rd Annu. Hawaii Int. Conf. Syst. Sci.*, vol. 0, no. c, pp. 1–5, 2000.

[11] R. Weron, "Electricity price forecasting: A review of the state-of-the-art with a look into the future," *Int. J. Forecast.*, vol. 30, no. 4, pp. 1030–1081, 2014.

[12] M. Sahni, R. Jones, and Y. Cheng, "Beyond the Crystal Ball - Locational Marginal Price Forecasting and Predictive Operations in the U.S. Power Markets," *IEEE Power and Energy Magazine*, no. 10(4), pp. 35–42, 2012.

[13] P. Cramton, "Electricity market design," *Oxford Rev. Econ. Policy*, vol. 33, no. 4, pp. 589–612, 2017.

[14] Ofgem, "Wholesale Energy Markets in 2015," 2015.

[15] O. Alsaç *et al.*, "The rights to fight price volatility," *IEEE Power Energy Mag.*, vol. 2, no. 4, pp. 47–57, 2004.

[16] US Department of Energy, "National Electric Transmission Congestion Report," 2015.

[17] PJM, "PJM LMP Model Information," 2017. [Online]. Available: goo.gl/tyzPuD. [Accessed: 13-Sep-2017].

[18] Western Power Distribution, "WPD Network Capacity Map." [Online]. Available: goo.gl/xxrGex. [Accessed: 01-Aug-2017].

[19] APX GROUP, "EPEX Spot Price Dashboard." [Online]. Available: goo.gl/UUknMP. [Accessed: 14-Feb-2018].

[20] National Grid, "Electricity Ten Year Statement 2016," 2016.

[21] Western Power Distribution, "Long Term Development Statement for Western Power Distribution (South West) plc's Electricity Distribution System," 2016.

[22] L. Kane and G. Ault, "A review and analysis of renewable energy curtailment schemes and Principles of Access: Transitioning towards business as usual," *Energy Policy*, vol. 72, pp. 67–77, 2014.